\documentclass[conference]{IEEEtran}
\usepackage{amsmath,amsthm,amssymb}
\usepackage{graphicx}
\usepackage{mathtools}
\usepackage{epstopdf}
\input epsf
\usepackage{bm}

\begin{document}
\title{List-Based Detection and Selection of Access Points in  Cell-Free Massive MIMO Networks}

\author{Tonny Ssettumba, Roberto B. Di Renna, Lukas T. N. Landau and Rodrigo C. de Lamare
\thanks{Tonny Ssettumba, Roberto B. Di Renna, Lukas T. N. Landau and Rodrigo C. de Lamare, Center for Telecommunications Studies (CETUC), Pontifical Catholic University of Rio de Janeiro (PUC-Rio), E-mail: tssettumba@aluno.puc-rio.br, \{roberto.brauer, lukas.landau, delamare\}@cetuc.puc-rio.br }%
}

\maketitle

\begin{abstract}
This paper proposes a cell-free massive multiple- input
multiple-output (CF-mMIMO) architecture with joint list- based
detection with soft interference cancelation (soft-IC) and access
points (APs) selection. In particular, we derive a new closed-form
expression for the minimum mean-square error receive filter while
taking the uplink transmit powers and APs selection into account.
This is achieved by optimizing the receive combining vector by
minimizing the mean square error between the detected symbol
estimate and transmitted symbol, after canceling the multi-user
interference (MUI). By using low-density parity check (LDPC) codes,
an iterative detection and decoding (IDD) scheme based on a message
passing is devised. In order to perform joint detection at the
central processing unit (CPU), the access points locally estimate
the channel and send their received sample data to the CPU via the
front haul links. In order to enhance the system's bit error rate
performance, the detected symbols are iteratively exchanged between
the joint detector and the LDPC decoder in log likelihood ratio
form. Furthermore, we draw insights into the derived detector as the
number of IDD iterations increase. Finally, the proposed list
detector is compared with existing detection techniques.
\end{abstract}

\section{Introduction}

Unlike centralized massive multiple-input multiple-output (MIMO)
systems \cite{mmimo,wence}, cell-free massive MIMO (CF-mMIMO)
systems operate by deploying a relatively large number of either
single-antenna or multiple-antenna access points (APs) in a
distributed fashion. The aim is to increase the network throughput,
coverage, spectral efficiency, energy efficiency and quality of
service \cite{rr2}. The APs send their received data signals to a
central processing unit (CPU) for information processing and
detection. The CPU operates the system at a network level with the
aim of coherent transmission and reception without necessarily
requiring cell boundaries \cite{r9}. Earlier works on CF-mMIMO
considered system architectures where the UEs are equipped with
single antennas and served by multiple APs \cite{r9}. However, such
systems require a significant number of front haul links between the
APs and the CPU. More recent developments in the CF-mMIMO
architecture have considered APs selection strategies that are
capable of reducing the complexity in the system architecture as
well as yielding more practical implementations. These APs selection
methods are capable of achieving close to the entire network
performance but with an added advantage of reducing the signaling
overheads \cite{rr2,rr4,cfrmmseprec}. CF-mMIMO networks are liable
to multi-user interference (MUI) caused by pilot contamination as
well as the overlapping of the signals transmitted by the users
during uplink data transmission phase \cite{rr6}. MUI makes the
receiver design complex and thus calls for efficient techniques in
the design of CF-mMIMO receivers. The key aspect in the design of
efficient receivers lies at reducing the error between detected
symbol and transmitted symbol data \cite{r1,r3,spa,r5}.

The performance of CF-mMIMO receivers can be enhanced by error
correcting codes (ECC) such as low-density parity check (LDPC)
\cite{peg,memd} and turbo codes \cite{r1}. The use of iterative
detection and decoding (IDD) techniques has been extensively studied
to improve the performance of co-located MIMO (Col-MIMO) and massive
MIMO (mMIMO) \cite{r1,r3,spa,r5}.  IDD-based detection techniques
leverage on message passing by exchanging soft beliefs in terms of
log-likelihood ratios (LLRs) between the detector and the decoder.
LDPC codes are cost-effective and have been used in the state-of-the
art standards to improve the performance of MIMO systems
\cite{r5,r7,bfidd,r10}.

 In this work, we present a joint IDD scheme with APs selection assuming imperfect channel estimation and taking the UL transmission powers and joint detection at the CPU into account. Particularly,  we derive a new closed-form expression for the MMSE detector with soft interference cancellation (MMSE-soft-IC). Based on the available a-priori information about the expectation of the transmitted symbol estimate, we draw insights into the derived detector and present the MMSE filters for the uplink as the number of IDD iterations increase. Furthermore, we propose a list-based detector to reduce the error propagation that exists in the interference cancellation step.
The bit error (BER) performance of the proposed  list-based detector
and APs selection is compared with soft MMSE and MMSE-soft-IC
detection techniques, for the system with APs selection (APs-Sel)
and without APs selection (All-APs).

The rest of this paper is organized as follows: Section \ref{sys}
presents the  proposed centralized system model for the  CF-mMIMO
architecture, the channel estimation and APs selection criterion.
The derived receive filter analysis and insights are presented in
section \ref{RCA}.   The  proposed list-based detector is presented
in \ref{MFSIC}. Section \ref{IDD} discusses the IDD scheme.
Simulation results and discussions are presented in \ref{Num_Dis}.
Section \ref{CO_FD} gives the concluding remarks.

\textbf{Symbol notations}: Lower bold and upper bold letters are
used  to represent vectors and matrices, respectively. The Hermitian
transpose operator is denoted by $(\cdot)^{H}$,
$\mathbb{E}\left\{x\right\}$ denotes the expected value of random
variable $X$.

\section{Proposed System Model}
\label{sys}

\begin{figure}[htbp]
\centering
\includegraphics[width=8cm]{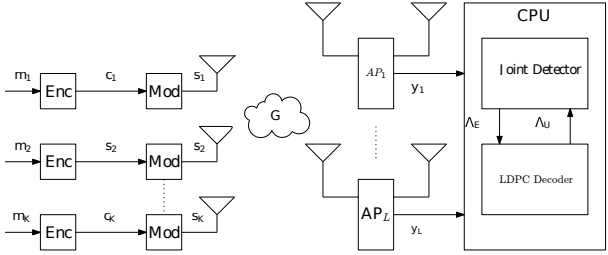}
\caption{Block diagram  for IDD scheme with Centralized Processing.}\label{fig1}
\end{figure}

We consider an uplink CF-mMIMO system model with imperfect channel
estimation. More specifically, an LDPC-coded  CF-mMIMO system
comprising of $L$ APs each equipped with $N$ receive antennas, $K$
single-antenna user equipment (UEs), a joint detector and an LDPC
decoder at the CPU for the considered  centralized processing
scenario  is considered as shown in Figure \ref{fig1}.
The data are first  encoded (Enc) by an  LDPC encoder having a code
rate ${R}$. This encoded sequence is then modulated (Mod) to complex
symbols with a complex  constellation of  $2^{M_{c}}$ possible
signal points and average energy $E_{s}$. The coded data is then
transmitted by ${K}$ UEs to the APs. During the data reception, the
APs act as relays and send the received information to the CPU which
comprises of a joint detector and an LDPC decoder. Then the joint
detector sends the received  soft information $\Lambda_{E}$ in the
form of LLRs to the LDPC decoder.   The decoder adopts an iterative
strategy by sending extrinsic information $\Lambda_{U}$ to the joint
detector which improves the performance of the entire network.
Additionally, the  performance of the proposed detector is examined
for the case with and without iterations.
 \subsection{Uplink Pilot Transmission and Channel Estimation}
 We start by assuming $\tau_{p}-$length pilot mutually orthogonal signals $\psi_{1}$, ..., $\psi_{\tau_{p}}$ with $\psi_{t}\psi^{*}_{t} = 1$ are used to estimate the channel. Furthermore, we assume that $K>\tau_{p}$ such that more than one UE can be assigned per pilot. The index of UE $k$ that uses the same pilot is denoted as $t_{k} \in \left\{1,..., \tau_{p}\right\}$ with $\vartheta_{k} \subset\left\{1, ..., K\right\}$ as the subset of UEs that use the same pilot as UE $k$ inclusive. The received $\mathbf{R}_{l}$ complex $N\times\tau_{p}$ signal after the UE transmission \cite{rr2} is given by \begin{align}
     \mathbf{R}_{l}=\sum_{j=1}^{K}\sqrt{\eta_{j}}\mathbf{g}_{jl}\bm{\psi}^{T}_{t_{j}}+\mathbf{N}_{l},
 \end{align}
 where $\eta_{j}$ is the transmit power from UE $j$, $\mathbf{N}_{l}$ is a receiver noise signal with independent ${\mathcal{N}}_{\mathbb{C}}\sim\left(0, \sigma^{2}\right)$ with noise power $\sigma^{2}$, $\mathbf{g}_{jl}\sim\mathcal{N}_{\mathbb{C}}\left(0, \bm{\Omega}_{jl}\right)$, and $\bm{\Omega}_{jl}\in\mathbb{C}^{NL\times NL}$ is the spatial correlation matrix that describes the channel's spatial properties between the $k$-th UE and $l$-th AP, $\beta_{k,l}\triangleq \frac{\mathrm{tr}\left(\bm{\Omega}_{jl}\right)}{N}$ is the large-scale (LS) fading coefficient. The AP first correlates the received signal with the associated normalized pilot signal $\bm{\psi}_{t_{k}}/\sqrt{\tau_{p}}$ to
$\mathbf{r}_{t_{kl}}\triangleq
\frac{1}{\sqrt{\tau_{p}}}\mathbf{R}_{l}\bm{\psi}^{*}_{t_{k}}\in
\mathbb{C}^{N}$ to estimate the channel $\mathbf{g}_{jl}$ given by
\begin{align}
  \mathbf{r}_{t_{kl}}=\sum_{j\in \vartheta_{k}}\sqrt{\eta_{j}\tau_{p}}\mathbf{g}_{jl}+\mathbf{n}_{t_{kl}},
\end{align}
where $\mathbf{n}_{t_{kl}}\triangleq
\frac{1}{\sqrt{\tau_{p}}}\mathbf{N}_{l}\bm{\psi}^{*}_{t_{k}}\sim\mathcal{N}_{c}\left(0,
\sigma^{2}\mathbf{I}_{N}\right)$ is the obtained noise sample after
estimation. Using \cite{rr2}, the MMSE estimate of $\mathbf{g}_{kl}$
is given by
\begin{align}
    \hat{\mathbf{g}}_{kl}=\sqrt{\eta_{k}\tau_{p}}\bm{\Omega}_{kl}\Psi^{-1}_{t_{k}}\mathbf{r}_{t_{kl}},
\end{align}
where $\Psi_{t_{kl}}=\mathbb{E}\left \{
\mathbf{r}_{t_{kl}}\mathbf{r}^{H}_{t_{kl}} \right \}=\sum_{j\in
\vartheta_{k}}{\eta_{j}\tau_{p}}\mathbf{\Omega}_{jl}+\mathbf{I}_{N}$
is the received signal vector correlation matrix. The channel
estimate $\hat{\mathbf{g}}_{kl}$ and estimation error
$\tilde{\mathbf{g}}_{kl}=\mathbf{g}_{kl}-\hat{\mathbf{g}}_{kl}$ are
independent with distributions
$\hat{\mathbf{g}}_{kl}\sim\mathcal{N}_{c}\left(0,
\eta_{k}\tau_{p}\bm{\Omega}_{kl}\mathbf{\Psi}^{-1}\bm{\Omega}_{kl}\right)$
and $\tilde{\mathbf{g}}_{kl}\sim\mathcal{N}_{c}\left(0,
\mathbf{C}_{kl}\right)$, where the parameter $\mathbf{C}_{kl}$ is
given by
\begin{align}
    \mathbf{C}_{kl}=\mathbb{E}\left \{ \tilde{\mathbf{g}}_{kl} \tilde{\mathbf{g}}^{H}_{kl}\right \}=\bm{\Omega}_{kl}-\eta_{k}\tau_{p}\bm{\Omega}_{kl}\mathbf{\Psi}^{-1}\bm{\Omega}_{kl}.
\end{align}
Pilot-contamination is created by the mutual interference generated
by the UEs sharing the same pilot signals. This  degrades the
system's performance \cite{rr2}. The received signal vector at the
APs  is given by
 \begin{align}
     \mathbf{y}=\sum_{j=1}^{K}\mathbf{g}_{j}s_{j}+\mathbf{n}.
 \end{align}
 This can be given in a more compact representation as
 \begin{align}\label{receiv_CPU_signal}
     \mathbf{y}=\mathbf{G}\mathbf{s}+\mathbf{n},
 \end{align}
  where ${\mathbf{G}}$ $\in \mathbb{C}^{NL\times K}$ is the  channel matrix with both small scale and LS fading coefficients.  $\mathbf{s}=[s_{1},..,s_{k-1}, s_k, s_{k+1},...,s_{K}]$ denotes the transmit symbol vector with $\mathbb{E}\left\{s_{k}s^{*}_{k}\right\}=\rho_{k}$, $\bm{\rho}=\left [ \rho_{1},..,\rho_{K} \right ]^{T}$ is the average transmit power vector,  $\mathbf{n}$ is the additive white Gaussian noise sample (AWGN) with zero mean and unit variance.

\subsection{Access Point Selection Procedure}
The APs selection considers an improved dynamic cooperation
clustering (DCC) approach presented in \cite{rr2}. This is achieved
by forming a block diagonal matrix
$\mathbf{D}_{kl}=\mathrm{diag}\left(\mathbf{D}_{k1},..,\mathbf{D}_{kL}\right)\in
\mathbb{C}^{NL\times NL}$, where $k=1,..,K$ and $l=1,..,L$. The
matrix determines which APs antennas or AP for the case of
single-antenna APs is going to serve a particular UE. Then the set
of UEs served  by AP $l$ is given by
\begin{align}
    \mathcal{D}_{l}=\biggl\{k:\mathrm{tr}\left(\mathbf{D}_{kl}\right)\geq 1,k\in\left\{1,..,K\right\}\biggr\}.
\end{align}.
The DCC does not alter the received signal  because all APs
physically receive the broadcast signal. The key aspect is to only
have a set of selected APs to take part during signal detection. The
the received signal after AP selection is given by
 \begin{align}\label{receiv_CPU_signalAPSEL}
     \mathbf{y}=\mathbf{D}_{k}\mathbf{G}\mathbf{s}+\mathbf{D}_{k}\mathbf{n},
 \end{align}
where
$\mathbf{D}_{k}=\mathrm{diag}\left(\mathbf{D}_{k1},..,\mathbf{D}_{kL}\right)$
is a block diagonal matrix which determines which APs are serving a
given UE or set of UEs. A special case occurs when
$\mathbf{D_{k}}=\mathbf{I}_{NL}$. This implies that all the UEs are
served by  all APs, implying that \eqref{receiv_CPU_signalAPSEL}
reduces to \eqref{receiv_CPU_signal}. The choice of which AP(s)
participate in service of particular UE is based on the joint access
point selection algorithm presented in \cite{rr2}. In this case the
UE appoints a master AP that is used to coordinate UL detection and
decoding based on the largest large scale fading (LLSF) coefficient.
The CPU then sets threshold a value $\beta_{th}$ for other
non-master APs to participate in services of a particular UE.
Further details of this selection algorithm can be found in
\cite{rr2}.

\section{Proposed Receiver Design}\label{RCA}
In this section, we present the derivations of the proposed receive
filter and structure. The proposed detector is capable of canceling
the MUI that occurs due to the other $K-1$ UEs in the network. Thus,
we propose a detector that comprises an MMSE filter followed by a
soft interference canceler. The demodulator forms soft estimates of
the transmitted symbols by computing the symbol mean $\bar{s}_{j}$
based on the available soft  beliefs or a-priori information from
the LDPC decoder \cite{r3}. This symbol mean is key in the
cancellation step because it determines if we have a perfect
interference canceler or a conventional MMSE filter as the number of
iterations increase. The symbol mean $\mathbb{E}\left \{ s_{j}
\right \}= \bar{s_{j}}$ is  given by
\begin{equation}
\label{expectation_sym}
       \bar{s}_{j} = \sum_{s \in {A}} s P(s_{j}=s),
\end{equation}
where ${A}$ is the complex constellation set.  The variance of the
$j$-th user symbol is computed as
    \begin{align}
    \sigma^{2}_{j} =\sum_{s\in \mathcal{A}}|s-\bar{s}_{j}|^{2}P(s_{j}=s).
    \end{align} The a-priori probabilities obtained from the  extrinsic LLRs are given by
\begin{align}\label{aprior_prob}
     P(s_{j}=s)=\prod_{l=1}^{M_{c}}\lbrack 1+\exp(-s^{b_{l}}\Lambda_{c}(b_{(j-1)M_{c}+l}))\rbrack^{-1},
\end{align}
where $s^{b_{l}}\in (+1,-1)$ denotes the value of the $l$-th bit of
symbol $s$,  $\Lambda_{c}(b_{i})$  denotes the extrinsic LLR of the
$i$-th bit computed by the LDPC decoder in the previous iteration.
We define $\Lambda_{c}(b_{i})=0$ at the first iteration since the
only available belief is from the channel. The probabilities in
\eqref{aprior_prob} are obtained by assuming  statistical
independence of bits within the same symbol  \cite{r3}.  Next we
present the derivations of the proposed MMSE-soft-IC detector.

{\bf Centralized Processing With APs Selection}: The aim of the
centralized detection after APs selection is to avoid redundant
processing of poor quality signals. Also the number of front haul
links significantly reduce which makes the system more scalable
\cite{rr2}. This leads to a more efficient implementation of the
CF-mMIMO system as well as avoiding wastage of resources. The major
draw back of such a selection scheme is the slight reduction in
performance. Therefore, there is a trade off between performance and
hardware complexity in the design by using AP selection techniques.
The received signal after APs selection is given by
\begin{align}\label{rece_cp}
    &\mathbf{y}^{CP}=\\&\mathbf{D}_{k}\hat{\mathbf{g}}_{k}s_{k}+\mathbf{D}_{k}\hat{\mathbf{G}}_{\text{i}}\mathbf{s}_{\text{i}}+\sum_{m=1}^{K}\mathbf{D}_{k}\tilde{\mathbf{g}}_{m}s_{m}+\mathbf{D}_{k}\mathbf{n}\notag,
\end{align}
where $\mathbf{y}^{CP}$ is an $NL\times 1$ vector consisting of the
received signals after APs selection,  The parameters  $s_{k}$,
$\hat{\mathbf{g}}_{k}$,  are the    transmitted symbol,  $NL\times
1$ estimated channel vector for the $k-$th UE, respectively. The
parameters  $\mathbf{s}_{\text{i}}$,   $\hat{\mathbf{G}}_{\text{i}}$
denote  the $K-1\times 1$  transmitted symbol vector and the
$NL\times K-1$ estimated channel matrix for the other $K-1$ UEs,
respectively. The parameter  $s_{m}$ and $\tilde{\mathbf{g}}_{m}$
are the transmitted symbol vector during channel estimation and
channel estimation error vector, respectively.

The decision statistic $y_{k}$ of the $k$-th user stream after
applying  the  receive combining vector $\mathbf{w}_{k}$ is given by
\begin{align}\label{eq2cp}
 & y^{CP}_{k}=\\&\mathbf{w}^{H}_{k}\biggl(\underbrace{\mathbf{D}_{k}\hat{\mathbf{g}}_{k}s_{k}}_{{\beta}_{k}}+\underbrace{\mathbf{D}_{k}\hat{\mathbf{G}}_{\text{i}}\mathbf{s}_{\text{i}}}_{\lambda_{\text{i}}}+\underbrace{\sum_{m=1}^{K}\mathbf{D}_{k}\tilde{\mathbf{g}}_{m}s_{m}}_{\alpha_{\text{e}}}+\underbrace{\mathbf{D}_{k}\mathbf{n}}_{\gamma_{n} }\biggr), \notag
\end{align}
where $\mathbf{\beta}_{k}$,  ${\lambda_{\text{i}}}$,
$\alpha_{\text{e}}$,  $\gamma_{k}$  denote  the desired signal for
$k$-th UE, the MUI from the other $K-1$ UEs, the channel estimation
error term and the phase-rotated noise. The estimated detected
symbol at the CPU  after removing the MUI is  given by
\begin{align}\label{funAPS1}
    \tilde{s}_{k}=\mathbf{w}^{H}_{k}\mathbf{y}^{CP}-\mathbf{w}^{H}_{k}\mathbf{D}_{k}\hat{\mathbf{G}}_\text{i}\bar{\mathbf{s}}_{\text{i}}.
\end{align}
 The  optimization of the receive combining  vector $\mathbf{w}_{k}$ enables us to minimize  the  mean square error in the detected data-stream.  We follow similar procedures in \cite{r1,r11} by optimizing the linear detection vector $\mathbf{w}_{k}$. The optimization problem is formulated as:
\begin{align}\label{funp2}
   \mathbf{w}_{k}   =\mathsf{arg}\min_{\left ( \mathbf{w}_{k}\right )}\mathbb{E}\biggl\{||\tilde{s}_{k}-s_{k}||^{2}\mid \hat{\mathbf{G}}\biggr\}.
\end{align}
The term $\mathbb{E}\left\{||\tilde{s}_{k}-s_{k}||^{2}\mid
\hat{\mathbf{G}}\right\}$ in \eqref{funp2} is given on the top of
the next page by \eqref{funp21}.
\begin{figure*}
   \noindent{\rule{\linewidth}{0.4pt}}
\begin{align}\label{funp21}
 \mathbb{E}\biggl\{||\tilde{s}_{k}-s_{k}||^{2}\mid \hat{\mathbf{G}}\biggr\}=&\mathbf{w}^{H}_{k}\biggl (\mathbf{D}_{k}\biggl(\rho_{k}\hat{\mathbf{g}}_{k}\hat{\mathbf{g}}^{H}_{k}+\hat{\mathbf{G}}_\text{i}\mathbf{\Delta}_\text{i}\hat{\mathbf{G}}^{H}_\text{i}\biggr)\mathbf{D}^{H}_{k}+\mathbf{D}_{k}\biggl(\sum_{m=1}^{K}\left(\mid\bar{s}_{m}\mid^2+\sigma^{2}_{m}\right)\mathbf{C}_{m}+\sigma^{2}\mathbf{{I}}_{NL}\biggr)\mathbf{D}^{H}_{k}\biggr )\mathbf{w}_{k}\notag\\&-\rho_{k}\mathbf{w}^{H}_{k} \mathbf{D}_{k}\hat{\mathbf{g}}_{k} -\rho^{*}_{k} \hat{\mathbf{g}}^{H}_{k}\mathbf{D}^{H}_{k}\mathbf{w}_{k}+\rho_{k}.
\end{align}
   \noindent{\rule{\linewidth}{0.4pt}}
\end{figure*}
Where $\mathbf{C}_{m}$ is the cross-correlation matrix from channel
estimation  error of the $m$-th UE, the matrix $
\mathbf{\Delta}_{i}=\mathsf{diag}\left[{\sigma_{{1}}^{2}},...,{\sigma_{{k-1}}^{2}},
{\sigma_{{k+1}}^{2}},...,{\sigma_{{K}}^{2}}\right]$ in \eqref{det_2}
is highly relevant in the IDD scheme during the soft-IC step.
Differentiating \eqref{funp21} with respect to (w.r.t)
$\mathbf{w}^{*}_{k}$ and equating to zero yields
\begin{align}\label{det_2}
  &\mathbf{w}^{\mathrm{APS-Sel}}_{k}= \rho_{k}\biggl (\mathbf{D}_{k}\biggl(\rho_{k}\hat{\mathbf{g}}_{k}\hat{\mathbf{g}}^{H}_{k}+\hat{\mathbf{G}}_\text{i}\mathbf{\Delta}_\text{i}\hat{\mathbf{G}}^{H}_\text{i}\biggr)\mathbf{D}^{H}_{k}\notag\\&+\mathbf{D}_{k}\biggl(\sigma^{2}\mathbf{{I}}_{NL}+\sum_{m=1}^{K}\left(\mid\bar{s}_{m}\mid^2+\sigma^{2}_{m}\right)\mathbf{C}_{m}\biggr)\mathbf{D}^{H}_{k}\biggr )^{-1}\notag\\&\times\mathbf{D}_{k}\hat{\mathbf{g}}_{k},
\end{align}

\subsubsection{Insights into the obtained detector}

 The MMSE- soft-IC detector for the scenario that uses all APs  can be obtain from \eqref{det_2} by taking a special case when $\mathbf{D}_{k}=\mathbf{I}_{NL}$, and is given by
\begin{align}
     & \mathbf{w}^{\mathrm{All-APs}}_{k}=\rho_{k}\biggl (\rho_{k}\hat{\mathbf{g}}_{k}\hat{\mathbf{g}}^{H}_{k}+\hat{\mathbf{G}}_\text{i}\bm{\Delta}_\text{i}\hat{\mathbf{G}}^{H}_\text{i}\notag\\&+\sigma^{2}\mathbf{{I}}_{NL}+\sum_{m=1}^{K}\left(\mid\bar{s}_{m}\mid^2+\sigma^{2}_{m}\right)\mathbf{C}_{m}\biggr )^{-1}\hat{\mathbf{g}}_{k}.
\end{align}
 For the first iteration, $\bar{\mathbf{s}}_{\text{i}}=\mathbf{0}$ in \eqref{expectation_sym}. In this case we have a linear MMSE filter and the detected signal in \eqref{funAPS1} is given by
\begin{align}\label{MMSE}
&\tilde{s}_{k}=\rho_{k}\hat{\mathbf{g}}^{H}_{k}\mathbf{D}^{H}_{k}\biggl(\rho_{k}\mathbf{D}_{k}\hat{\mathbf{g}}_{k}\hat{\mathbf{g}}^{H}_{k}\mathbf{D}^{H}_{k}+\mathbf{D}_{k}\hat{\mathbf{G}}_\text{i}~\mathrm{diag}\left(\bm{\rho}_\text{i}\right)\hat{\mathbf{G}}^{H}_\text{i}\mathbf{D}^{H}_{k}\notag\\&+\mathbf{D}_{k}\left(\sigma^{2}\mathbf{I}_{NL}+\sum_{m=1}^{K}\rho_{m}\mathbf{C}_{m}\right)\mathbf{D}^{H}_{k}\biggr
)^{-1}\mathbf{y}^{CP},
\end{align}
 where $\bm{\rho}_{i}$ denotes the average transmit power vector for the other $K-1$ UEs.
 As the number of iterations increases, $\bar{\mathbf{s}}_{\text{i}}\approx\mathbf{s}_{\text{i}}$ in \eqref{expectation_sym}. In such a scenario, the filter becomes a perfect interference canceler and thus \eqref{funAPS1} yields
\begin{align}\label{eq20}
&\tilde{s}_{k}=\rho_{k}\hat{\mathbf{g}}^{H}_{k}\mathbf{D}^{H}_{k}\biggl
(\rho_{k}
\mathbf{D}_{k}\hat{\mathbf{g}}_{k}\hat{\mathbf{g}}^{H}_{k}\mathbf{D}^{H}_{k}+\mathbf{D}_{k}\biggl(\sigma^{2}\mathbf{I}_{NL}\notag\\&+\sum_{m=1}^{K}\mid
s_{m}\mid^{2}\mathbf{C}_{m}\biggr)\mathbf{D}^{H}_{k} \biggr
)^{-1}\left (
\mathbf{y}^{CP}-\mathbf{D}_{k}\hat{\mathbf{G}}_\text{i}\mathbf{s}_\text{i}
\right ).
\end{align}

 \section{List-based detector}\label{MFSIC}

\begin{figure}[!h]
\centering
\includegraphics[width=9.5cm]{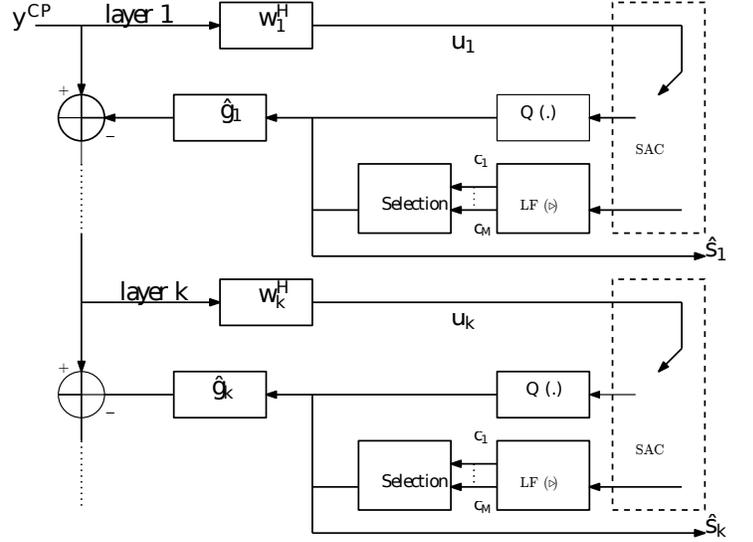}
\caption{Block diagram of the proposed list-based detector.}\label{fig11}
\end{figure}

In this section, we describe the operation of the proposed
list-based detection scheme shown in Fig.\ref{fig11}, which has been
inspired by the works in \cite{spa,r5,dfcc,mbdf,list_mtc}.

The design takes advantage of list feedback (LF) diversity by
selecting a list of constellation candidates if there is
unreliability of the previously detected symbols \cite{r5}. In this
case, a shadow area constraint (SAC) is initiated in order to obtain
an optimal feedback candidate. The SAC is capable of reducing search
space from growing exponentially as well as reducing the
computational complexity. The key idea of such a selection criterion
is to avoid redundant processing when there is a reliable decision.
The procedure of obtaining the detected symbol $\hat{s}_{k}$ of the
$k$-th user is analogous to the steps presented in \cite{r5}. The
$k$-th user soft estimate is obtained by
$u_{k}=\mathbf{w}_{k}^{H}\check{\mathbf{y}}_{k}$. The filter
$\mathbf{w}_{k}$ is similar to the receive MMSE filter in
\eqref{MMSE} and
$\check{\mathbf{y}}_{k}=\mathbf{y}^{CP}-\sum_{t=1}^{k-1}\mathbf{D}_{k}\hat{\mathbf{g}}_{t}\hat{{s}}_{t}$
represents the received vector following the soft cancellation of
the $k-1$ symbols that were previously detected. The SAC assesses
the reliability of this decision using the soft estimate $u_{k}$ for
each layer according to
\begin{align}
    d_{k}=\vert u_{k}-\nu_{f}\vert,
\end{align}
where $\nu_{f}=\mathsf{arg}\min_{\mathbf{\nu_{f}}\in{\mathcal{A}}}
\left\{ \vert u_{k}-\nu_{f}\vert\right\}$ denotes the closest
constellation point to the $k$-th user soft estimate $u_{k}$. If
$d_{k}>d_{\text{th}}$ the chosen constellation point gets dumped
into the shadow area of the constellation map since the choice is
deemed to be unreliable.  Parameter $d_{\text{th}}$ is the
predefined  threshold euclidean distance to guarantee reliability of
the selected symbol \cite{r5}. The list-based algorithm performs a
hard slice for UE $k$ as in the soft-IC if there is reliability of
the soft estimate $u_{k}$. In this case, $\hat{s}_{k}=Q(u_{k})$ is
the estimated symbol, where $Q(\cdot)$ is the quantization notation
which maps to the constellation symbol closest to  $u_{k}$.

Otherwise, the decision is deemed unreliable. In this case, a list
of candidates  $\mathcal{L}=\{c_{1},
c_{2},...,c_{m},...,c_{M}\}\subseteq\mathcal{A}$ is generated, which
is made up of the constellation $M$ points that are closest to $u_{
k}$. The number of candidate points $M$ is given by the QPSK
symbols. The algorithm selects an optimal candidate
$c_{m,\text{opt}}$ from a list of $\mathcal{L}$ candidates. Thus,
the unreliable choice $Q(u_{k})$ is replaced by a hard decision and
$\hat{s}_{k}=c_{m,\text{opt}}$ is obtained. The list-based detector
first defines the selection vectors
$\bm{\phi}^{1},\bm{\phi}^{2},...,\bm{\phi}^{m},...\bm{\phi}^{M}$
whose size is equal to the number of the constellation candidates
that are used every time a decision is considered unreliable. For
example, for the $k$-th  layer,  a $K\times 1$ vector $\bm{\phi}^{m}
   =\left [ \hat{s}_{1},...,\hat{s}_{k-1}, c_{m},\phi^{m}_{k+1},...,\phi^{m}_{q},...,\phi^{m}_{K} \right ]^T$  which is  a potential choice corresponding to $c_{m}$ in the k-th user comprise the following items: (a) The previously estimated symbols $\hat{s}_{1},\hat{s}_{2}, ..., \hat{s}_{k-1}$. (b) The candidate symbol $c_{m}$ obtained from the constellation for subtracting a decision that was considered unreliable $Q(u_{k})$ of the k-th user. (c) Using (a) and (b) as the previous decisions, detection of the next user data $k+1, ...,q,...,$ $K$-th is performed by the  soft-IC approach. Mathematically, the choice $\phi^{m}$ is given by \cite{r5}
\begin{align}
   \phi^{m}_{q}=Q(\mathbf{w}^{H}_{q}{\hat{\bm{y}}}^{m}_{q}),
\end{align}
where the index $q$ denotes a given UE between the $k+1$-th and the
$K$-th UE, $
\hat{\bm{y}}^{m}_{q}=\check{\bm{y}}_{k}-\mathbf{D}_{k}\hat{\mathbf{g}}_{k}c_{m}-\mathbf{D}_{k}
\sum_{p=k+1}^{q-1}\hat{\mathbf{g}}_{p}\phi^{m}_{p}$. A key attribute
of the list-based detector is that the same MMSE filter
$\mathbf{w}_{k}$ is used for all the constellation candidates.
Therefore, it has the same computational cost as the conventional
soft-IC. The optimal candidate ${m,\text{opt}}$ is selected
according to the local maximum likelihood (ML) rule given by
\begin{align}
  {m,\text{opt}}=\mathsf{arg}\min_{1\leq m\leq M}\left \|\bm{y}^{CP}-\mathbf{D}_{k}\hat{\mathbf{G}}\bm{\phi}^{m}\right\|^{2}.
\end{align}
\section{Iterative detection and decoding}\label{IDD}
In this section, the MMSE- based  detectors are presented for the
IDD scheme as shown  in Fig. \ref{fig1}, consisting of a joint
detector and an LDPC decoder. The received signal at the output of
the filter, contains the desired symbol, MUI, the channel estimation
error  and noise. We use similar assumptions given  in
\cite{r1,r3,r10} to approximate  the parameter $u_{k}$ as an AWGN
channel given by
\begin{align}
  u_{k}=\omega_{k}s_{k}+z_{k},
\end{align}
where the parameter $\omega_{k}$ is given by
$\mathbb{E}\{{s}^{*}_{k}u_{k}\}=\rho_{k}\mathbf{w}^{H}_{k}\mathbf{D}_{k}\hat{\mathbf{g}}_{k}$.
The parameter $z_{k}$ is a zero-mean AWGN variable. Using similar
procedures as in \cite{r5},  variance of $z_{k}$ is given by
$\kappa^{2}_{k}=\mathbb{E}\left\{\mid
u_{k}-\omega_{k}s_{k}\mid^{2}\right\}=\mathbf{w}^{H}_{k}\mathbf{D}_{k}\left
( \sum_{m=1}^{K}\rho_{m}\mathbf{C}_{m}+\sigma^{2}\mathbf{I}_{NL}
\right )\mathbf{D}^{H}_{k}\mathbf{w}_{k}$. The extrinsic LLR
computed by the detector for the $l$-th bit
$l\in\left\{1,2,...,M_{c}\right\}$ of the symbol $s_{k}$ transmitted
by the $k$-th user is \cite{r1,r3}
\begin{align}
  \Lambda_{E}\left ( b_{(k-1)M_{c}+l} \right )&=\log\frac{\sum _{s\in A^{+1}_{l}}f\left ( u_{k}|s \right )P\left (s \right )}{\sum _{s\in A^{-1}_{l}}f\left ( u_{k}|s \right )P\left (s \right )}\\&\notag-\Lambda_{U}\left ( b_{(k-1)M_{c}+l} \right ),
\end{align}
    where $A^{+1}_{l}$ is the set of $2^{Mc-1}$ hypothesis $s$ for which the $l$-th bit is $+1$. The a-priori probability $P(s)$ is given by \eqref{aprior_prob}. The approximation of the likelihood function  \cite{r3} $f(u_{k}|s)$ is given by
    \begin{align}
        f\left ( u_{k}|s \right )\simeq\frac{1}{\pi\kappa^{2}_{k}}\exp\left (-\frac{1}{\kappa^{2}_{k}} |u_{k}-\omega_{k}s|^{2} \right ).
    \end{align}
%\subsection
{\bf Decoding Algorithm}: The soft beliefs  are  exchanged between
the proposed detectors and the decoder in an iterative manner. The
traditional  sum product algorithm (SPA) suffers from performance
degradation caused by the tangent function especially in the
error-rate floor region \cite{r10}. Therefore, we use the box-plus
SPA in this paper because it yields less complex approximations. The
decoder is made up of two stages namely: The single parity check
(SPC) stage and the repetition stage. The LLR sent from check node
$(CN)_{J}$ to variable node $(VN)_{i}$ is computed as
\begin{align}
    \Lambda_{j\longrightarrow i}=\boxplus{i^{'}\in N(j)\diagdown i\Lambda_{i^{'\longrightarrow j}}}.
\end{align}
As shorthand, we use  $\Lambda_{1}\boxplus \Lambda_{2}$ to denote
the computation of $\Lambda(\Lambda_{1}\bigoplus \Lambda_{2})$. The
LLR is computed  by
\begin{align}
\Lambda_{1}\boxplus \Lambda_{2}=&\log\left (
\frac{1+e^{\Lambda_{1}+\Lambda_{2}}}{e^{\Lambda_{1}}+e^{\Lambda_{2}}}
\right ),\\\notag
  =&\mathrm{sign}(\Lambda_{1})\mathrm{sign}(\Lambda_{2})\min(\left | \Lambda_{1} \right |,\left | \Lambda_{2} \right |)\\\notag&+\log\left ( 1+e^{-\left |\Lambda_{1}+\Lambda_{2}  \right |} \right )-\log\left (1+e^{-\left |\Lambda_{1}-\Lambda_{2}  \right |}  \right ).
\end{align}
The LLR from $VN_{i}$ to $CN_{j}$ is given by
\begin{align}
   \Lambda_{i\longrightarrow j}=\Lambda_{i}+\sum_{j^{'}\in N(i)\backslash j}\Lambda_{j^{'}\longrightarrow i},
\end{align}
where the parameter $\Lambda_{i}$ denotes the LLR at $VN_{i}$,
${j^{'}\in N(i)\backslash j}$ denotes all CNs connected to $VN_{i}$
except $CN_{j}$. The exchange of LLRs can be further refined with
several strategies \cite{vfap}.

\section{Simulation results and discussion} \label{Num_Dis}

In this section, the BER performance of the  proposed soft detectors
is presented for the CF-mMIMO  and COL-mMIMO settings. The CF-mMIMO
channel exhibits high pathloss (PL) values due to  LS fading
coefficients. The SNR definition is given by
\begin{align}
    SNR=\frac{{tr}(\mathbf{G}~\mathrm{diag}\left(\bm{\rho}\right)\mathbf{G}^{H})}{\sigma^{2} NL K},
\end{align}
  The simulation parameters  are varied  as follows: We consider a cell-free environment with  a square of dimensions $D\times D=$, where $D=1$ km. respectively. The APs are deployed $10~m$ above the UE.  Bandwidth$=20$ MHz, $N=1$,  $d_{th}=0.38$, $\tau_{u}=190$, $\tau_{p}=10$, $\tau_{c}=200$, $\eta_{k}=100$ mW, the spatial correlation matrices $\bm{\Omega}_{jl}$ are assumed to be  locally available at the APs \cite{rr2}. We use an LDPC code with code word length  $C_{\text{leng}}=256$ bits, $M=128$ parity check bits and $C_{\text{leng}}-M$ message bits, the threshold for non-master AP to serve is set at $\beta_{\text{th}}=-60~\mathrm{dB}$ and the code rate $R=\frac{1}{2}$. The maximum number of inner iterations  (decoder iterations) is set to $10$.  The signal power $\rho=1$ W and the simulations are run for $10^{3}$ channel realizations. The modulation scheme used is quadrature phase shift keying (QPSK). The LS fading coefficients are obtained according to the 3GPP Urban Microcell model in \cite{rr2} given by $
    \beta_{k,l}\left[\mathbf{\mathrm{dB}}\right]=-30.5-36.7\log_{10}\biggl(\frac{d_{kl}}{1 m}\biggr)+\Upsilon_{kl},
$ where $d_{kl}$ is the distance between the $k$-th UE and $l$-th
AP,  $\Upsilon_{kl}\sim\mathcal{N}\left(0, 4^{2}\right)$ is the
shadow fading.

\begin{figure}[!ht]
\centering
\includegraphics[width=.9\linewidth]{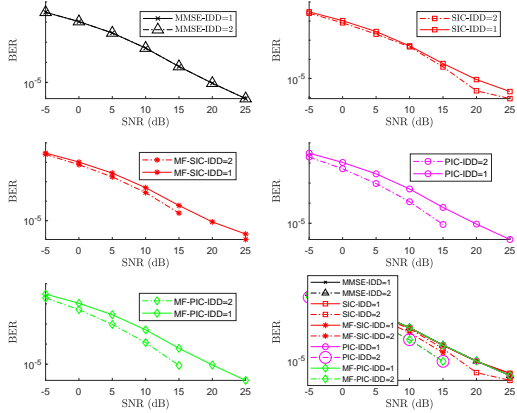}
\caption{BER versus SNR for All APs and Centralized AP Selection
varying number of IDD iterations   with $L=32$, $K=8$} %,  (a) MMSE,
%(b) MMSE-Soft-IC, (c) List-MMSE-Soft-IC.}
\label{fig4}
\end{figure}

\begin{figure}[!ht]
\centering
\includegraphics[width=.9\linewidth]{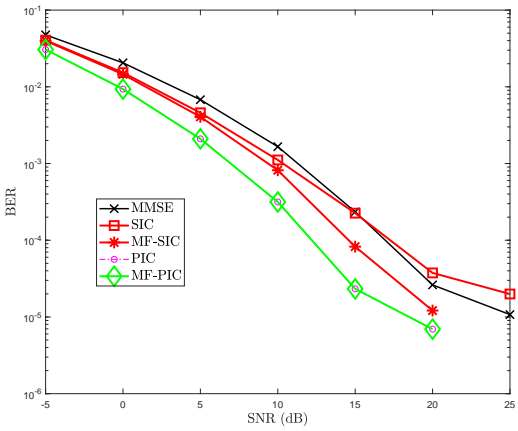}
\caption{BER versus SNR for All APs and Centralized AP Selection
with $L=32$, $K=8$, $\mathrm{IDD}=2$.} \label{fig5}
\end{figure}

Figure \ref{fig4} presents BER versus SNR as the number of IDD
iterations are varied for  (a) MMSE,(b) MMSE-Soft-IC and (c)
List-MMSE-Soft-IC detection schemes. It can be observed that
increasing number of iterations reduces the BER.  This is because
more a-posterior information is exchanged between the joint detector
and decoder as the iterations increase, which  improves the system
performance. For the case of MMSE, the number of iterations do not
reduce BER  because there is no $\bm{\Delta}_{\text{i}}$ in this
filter which is needed for the IDD scheme to improve the
performance. Figure \ref{fig5} presents the BER versus SNR for the
case with APs selection and case when using all the APs while
comparing the studied detectors. From this figure, It can be
observed that the List-MMSE-Soft-IC detector achieves lower BER
values, followed by the MMSE-Soft-IC detector and lastly the soft
MMSE detector. Also the case with APs selection achieves slightly
higher BER rates as compared to the case when all APs are used. This
is due to reduction in diversity order.

\section{Concluding Remarks}\label{CO_FD}
A joint list-based detector for CF-mMIMO architecture with
centralized APs selection and processing is  presented.
Specifically,  a new closed-form expression for the MMSE-soft-IC
detection scheme with APs selection is derived. The resulting MMSE
detectors are provided based on knowledge of the expectation of the
transmitted symbols. More particularly, a list-based detector that
can mitigate error propagation at the interference cancellation
stage and enhance BER performance is proposed. Additionally, the
soft MMSE and MMSE-Soft-IC detection techniques are contrasted with
the suggested list-based detector. The performance of a system with
and without APs selection is also  compared. The system that uses
all APs achieves lower BER values as compared to the one with APs
selection. Thus, there is a trade-off between scalability and BER
performance while using APs selection. The major advantage gained
with APs selection is the reduction in the signaling between the APs
and CPU which makes the network more scalable and practical.


\begin{thebibliography}{00}

\bibitem{mmimo}
R. C. de Lamare, "Massive MIMO systems: Signal processing challenges and future trends," in URSI Radio Science Bulletin, vol. 2013, no. 347, pp. 8-20, Dec. 2013.

\bibitem{wence}
W. Zhang et al., "Large-Scale Antenna Systems With UL/DL Hardware Mismatch: Achievable Rates Analysis and Calibration," in IEEE Transactions on Communications, vol. 63, no. 4, pp. 1216-1229, April 2015.

 \bibitem{rr2}
E.~Bj{\"o}rnson and L. Sanguinetti, '' Scalable Cell-Free Massive MIMO Systems'',  \emph{ IEEE Trans. Wireless Commun.},  vol. 68, no. 7, pp. 4247-4261, July 2020.

\bibitem{r9}
 H. Q. Ngo, A. Ashikhmin, H. Yang, E. G. Larsson and T. L. Marzetta, "Cell-Free Massive MIMO Versus Small Cells, " \emph{IEEE Trans. Commun.}, vol. 16, no. 3, pp. 1834-1850, March 2017.

 \bibitem{rr4}
H. T. Dao and S. Kim, '' Effective channel gain-based access point
selection in cell-free massive MIMO systems'',
\emph{ IEEE Access}, vol. 8, pp. 108127 - 108132, June 2020.

\bibitem{cfrmmseprec}
V. M. T. Palhares, A. R. Flores and R. C. de Lamare, "Robust MMSE Precoding and Power Allocation for Cell-Free Massive MIMO Systems," in IEEE Transactions on Vehicular Technology, vol. 70, no. 5, pp. 5115-5120, May 2021

 \bibitem{rr6}
Shakya, I.L., Ali, F.H.
        '' Joint access point selection and interference cancellation for cell-free massive MIMO'',
\emph{IEEE Commun. Lett.}, , vol. 25, no. 4, pp. 1313--1317, Apr. 2021.


\bibitem{peg}
Xiao-Yu Hu, E. Eleftheriou and D. M. Arnold, "Regular and irregular progressive edge-growth tanner graphs," in IEEE Transactions on Information Theory, vol. 51, no. 1, pp. 386-398, Jan. 2005.

\bibitem{memd}
C. T. Healy and R. C. de Lamare, "Design of LDPC Codes Based on Multipath EMD Strategies for Progressive Edge Growth," in IEEE Transactions on Communications, vol. 64, no. 8, pp. 3208-3219, Aug. 2016.

 \bibitem{r1}
 X. Wang and H.~V.~Poor, '' Iterative (turbo) soft interference cancellation and decoding for coded CDMA'',
\emph{ IEEE Trans. Commun.}, vol. 47, no. 7, pp. 1046--1061, Jul.1999.

\bibitem{r11}
 M.~ Sellathurai and S.~ Haykin, '' TURBO-BLAST for wireless communications: Theory and experiments'', \emph{ IEEE Trans. Signal Process.}, vol. 50, no. 10, pp. 2538–-2546, Oct. 2002.

\bibitem{bfidd}
A. G. D. Uchoa, C. T. Healy and R. C. de Lamare, "Iterative Detection and Decoding Algorithms for MIMO Systems in Block-Fading Channels Using LDPC Codes," in IEEE Transactions on Vehicular Technology, vol. 65, no. 4, pp. 2735-2741, April 2016

\bibitem{r10}
Z.~Shao, R.~C.~de Lamare and L.~T.~N.~Landau, "Iterative Detection and Decoding for Large-Scale Multiple-Antenna Systems With 1-Bit ADCs, "\emph{ IEEE Wireless Commun. Lett.}, vol.~7, no.~3, pp.~476--479, Jun. 2018.

\bibitem{r3}
A.~Matache, C.~ Jones and R.~D.~Wesel, ''Reduced complexity MIMO detectors for LDPC coded systems'',  \emph{in~Proc.  IEEE Military Commun. Conf.}, Monterey, CA, USA, pp. 1073-1079, 31 Oct.-3 Nov. 2004.

\bibitem{spa}
R. C. De Lamare and R. Sampaio-Neto, "Minimum Mean-Squared Error Iterative Successive Parallel Arbitrated Decision Feedback Detectors for DS-CDMA Systems," \emph{IEEE Trans. Commun.}, vol. 56, no. 5, pp. 778--789, May 2008.

\bibitem{r5}  P.~Li, R.~C.~de Lamare and R.~Fa,
 ''Multiple Feedback Successive Interference Cancellation Detection for Multiuser MIMO Systems'', \emph{IEEE Trans. Commun.}, vol. 10, no. 8, pp. 2434--2439, Jun. 2011.

\bibitem{dfcc}
P. Li and R. C. De Lamare, "Adaptive Decision-Feedback Detection With Constellation Constraints for MIMO Systems," in IEEE Transactions on Vehicular Technology, vol. 61, no. 2, pp. 853-859, Feb. 2012

\bibitem{mbdf}
 R. C. de Lamare, "Adaptive and Iterative Multi-Branch MMSE Decision Feedback Detection Algorithms for Multi-Antenna Systems," in IEEE Transactions on Wireless Communications, vol. 12, no. 10, pp. 5294-5308, October 2013.

\bibitem{did}
P. Li and R. C. de Lamare, "Distributed Iterative Detection With Reduced Message Passing for Networked MIMO Cellular Systems," in IEEE Transactions on Vehicular Technology, vol. 63, no. 6, pp. 2947-2954, July 2014.

\bibitem{list_mtc}
R. B. Di Renna and R. C. de Lamare, "Iterative List Detection and Decoding for Massive Machine-Type Communications," in IEEE Transactions on Communications, vol. 68, no. 10, pp. 6276-6288, Oct. 2020.


\bibitem{vfap}
J. Liu and R. C. de Lamare, "Low-Latency Reweighted Belief Propagation Decoding for LDPC Codes," in IEEE Communications Letters, vol. 16, no. 10, pp. 1660-1663, October 2012


\bibitem{r7}
 C. D’Andrea and E. G. Larsson, "Improving Cell-Free Massive MIMO by Local Per-Bit Soft Detection, " \emph{ IEEE Commun. Lett.}, vol. 25, no. 7, pp. 2400--2404, Apr. 2021.


\end{thebibliography}
\end{document}